\documentclass{article}

\usepackage{color}

\def\giorno{6/10/2014}

\def\a{\alpha}
\def\b{\beta}

\def\ga{\gamma}
\def\de{\delta}   
\def\eps{\varepsilon}
\def\phi{\varphi}
\def\la{\lambda}

\def\s{\sigma}

\def\om{\omega}

\def\D{{\mathcal D}}

\def\G{{\mathcal G}}

\def\I{{\mathcal I}}
\def\P{{\mathcal P}}
\def\R{{\bf R}}

\def\X{{\mathcal X}}
\def\Y{{\mathcal Y}}
\def\T{{\mathrm T}}

\def\Ga{\Gamma}
\def\De{\Delta}
\def\La{\Lambda}

\def\pa{\partial}

\def\d{{\rm d}}       

\def\ss{\subset}

\def\EOR{\hfill {$\odot$}}
\def\EOP{\hfill {$\triangle$}}

\def\({\left(}
\def\){\right)}
\def\[{\left[}
\def\]{\right]}
\def\=#1{\overline #1}

\def\~#1{\widetilde #1}
\def\wt#1{\widetilde #1}
\def\.#1{\dot #1}
\def\^#1{\widehat #1}

\def\mapright#1{\smash{\mathop{\longrightarrow}\limits^{#1}}}
\def\mapdown#1{\Big\downarrow\rlap{$\vcenter{\hbox{$\scriptstyle#1$}}$}}

\def\interno{\hskip 2pt \vbox{\hbox{\vbox to .18
truecm{\vfill\hbox to .25 truecm
{\hfill\hfill}\vfill}\vrule}\hrule}\hskip 2 pt}

\def\beq{\begin{equation}}
\def\eeq{\end{equation}}
\def\eqref#1{(\ref{#1})}

\def\symmref{EMS1,CGbook,KrV,Olv1,Olv2,Ste}

\begin{document}

\title{Symmetry and Lie-Frobenius reduction of differential equations}

\author{G. Gaeta\thanks{Research partially supported by MIUR-PRIN program
under project 2010-JJ4KPA} \\
{\it Dipartimento di Matematica, Universit\`a degli Studi di Milano,} \\
{\it via C. Saldini 50, 20133 Milano (Italy)} \\ {\tt
giuseppe.gaeta@unimi.it}}

\date{Revised version \ -- \ \giorno}

\maketitle

\begin{abstract}
\noindent
Twisted symmetries, widely studied in the last decade,
proved to be as effective as standard ones in the analysis and
reduction of nonlinear equations. We explain this effectiveness in
terms of a Lie-Frobenius reduction; this requires to focus not
just on the prolonged (symmetry) vector fields but on the
distributions spanned by these and on systems of vector fields in
involution in Frobenius sense, not necessarily spanning a Lie
algebra.

\end{abstract}

\section*{Introduction and motivation}
\label{sec:intro}

Nonlinear systems are relevant in Mathematics and Physics, but
they are as a rule hard to analyze. One of the key tools in
tackling nonlinear differential equations (by this we always mean
possibly a system) is provided by symmetry analysis
\cite{\symmref}, and actually this tool -- and more generally the
theory of Lie groups -- was created by S. Lie precisely to study
nonlinear differential equations.

The theory received a sound geometrical setting thanks to the work
of Cartan and Ehresmann, with the theory of Jet bundles. It
received a new boost several decades after the work of Lie and
Cartan, when the work of Ovsjannikov in the USSR and by Birkhoff
in the USA \cite{Bir,Ovs} revived interest in it. Applications of
the theory beyond the simplest case require rather extensive
computations, but these are nowadays standard thanks to computer
algebraic manipulation languages \cite{HeW1,HeW2,HeW3,Sch}.

The renewal of interest also called for generalizations of the
theory; thus different kind of symmetries extending the classical
concept were considered in the literature. In all of these cases,
one considers vector fields in the extended phase manifold (which
later on in this paper will be denoted as $M$) and their standard
prolongation to the relevant Jet bundle $J^n M$; generalization
consists either in considering vector fields more general than
Lie-point ones (e.g. generalized vector fields \cite{KrV,Olv1}), or
in weakening the requirements regarding their action on the set of
solutions of the equation under study (e.g. requiring that there
are some invariant solutions albeit the vector fields do not fully qualify as
symmetries, as in conditional or nonclassical symmetries
\cite{ClK,LeWi1,LeWi2}; or that only a subset of solutions is
invariant under their action, as in partial symmetries
\cite{CGpart}). The fact that one always considers standard
prolongation is entirely natural, as once the transformation of
independent and dependent variables are assigned, the
transformation laws for derivatives of the latter with respect to
the former are also entirely determined: this is precisely the
content of the standard prolongation formula \cite{\symmref}.

In recent years, starting with the work of Muriel and Romero in
2001, it was realized that one can extend the theory in a
different, and somehow surprising, direction. That is, it was
realized that one could consider a deformation, or a ``twisting'',
of the prolongation operation, having twisted prolongations of
Lie-point vector fields (or of more general types of vector
fields, but here we will only focus on the Lie-point case). When
these twisted prolongation satisfy the usual relation with (the
geometrical object in $J^n M$ representing) the differential
equations under study, one speaks of \emph{twisted symmetries}.
The twisting is always based on an auxiliary object (a function, a
matrix, a one-form, depending on the type of twisted symmetry one
is considering; see below for details); when the latter vanishes,
one is reduced to standard prolongations and symmetries; thus,
twisted symmetries represent a genuine extension of standard ones.

The nontrivial fact is that these twisted symmetries turn out to
be \emph{as effective as standard ones} in the analysis
of differential equations. This applies both to ODEs, where one looks
for a symmetry reduction of the system under study, and to
PDEs, where one looks for invariant solutions. In fact, as twisted
symmetries are more general than standard ones, there are cases
where we have no standard symmetry but there are twisted
symmetries (as was the case in the pioneering study of Muriel and
Romero on $\la$-symmetries \cite{MuRom1,MuRom2}), so that
equations which cannot be reduced or integrated within the
framework of standard symmetries turn out to be integrated via the
standard procedure if one resorts to twisted symmetries.
In other words, twisted symmetries are an extra tool to study nonlinear equations; and one which can work where standard symmetries fail.

The purpose of this paper is to understand \emph{why} twisted symmetries
are as effective as standard ones. In order to do this, we will
slightly change the usual focus in determining the symmetry
properties of differential equations (with a formulation fully
equivalent to the standard one), looking -- even in the case of a
single vector field -- at \emph{distributions} rather than at vector
fields. This will immediately call for consideration of possible
changes of the set of generators for the distribution, and we will
find that such a change leads to considering twisted prolongations
and symmetries.

In view of the dominant role assumed by distributions, and by the
involution rather than the Lie algebraic structure, we will speak
of \emph{Lie-Frobenius reduction} to characterize our approach.

\bigskip\noindent
{\bf Acknowledgements.} My understanding of twisted symmetries is
based on common work with Giampaolo Cicogna, Paola Morando and
Sebastian Walcher; I am indebted to them. I also thank the referees
who helped improving the first version of this paper.
My research is partially supported by MIUR-PRIN program under project 2010-JJ4KPA.

\section{General framework}
\label{sec:frame}

Let us recall the general framework for the symmetry analysis of
differential equations. Our task here is mainly to fix notation,
and we will assume the reader has some familiarity with the
subject; see e.g. \cite{\symmref}. All the objects considered
(manifolds, functions, etc.) will be assumed to be smooth, i.e. of
class $C^\infty$, and real. We will use the Einstein convention
for sum on repeated indices and multi-indices.

\subsection{Notations for derivatives}

First of all, a word about compact notation for partial
derivatives. In the presence of coordinates $x^i$ and $u^a$, we
will write $$ \pa_i \ := \ \frac{\pa}{\pa x^i} \ , \ \ \pa_a \ :=
\ \frac{\pa}{\pa u^a} \ . $$ We will freely use the multi-index
notation for partial derivatives; thus while first order partial
derivatives of $u^a = u^a (x^1,...,x^q)$ will be denoted by $u^a_i
:= (\pa u^a / \pa x^i)$, for higher order derivatives we will
consider multi-indices $J = (j_1,...,j_q)$, with $j_i \in {\bf
N}$, of order $|J| = q_1 + ... + q_n$, and write $$ u^a_J \ := \
\frac{\pa^{|J|} u^a}{\pa (x^1)^{j_1} ... \pa (x^q)^{j_q} } \ . $$
The notation $\^J = \{J,i\}$ will denote the multi-index with
entries $\^j_k = j_k + \de_{ik}$, and correspondingly $u^a_{J,i}
:= (\pa u^a_J / \pa x^i)$. We will also need to consider
derivatives with respect to $u^a_J$; for these we will also use
the notation
$$ \pa_a^J \ := \ \frac{\pa}{\pa u^a_J} \ . $$
We will denote the set of derivatives (of the $u$ with respect to
the $x$) of order $n$ as $u^{[n]}$, that of derivatives of order
from $0$ to $n$ (included) as $u^{(n)}$; thus $f(x,u^{(n)})$ will
denote a function of $x$, $u$ and of derivatives of order up to
$n$.

Finally, we will consider total derivatives $D_i$ with respect to
the independent variables $x^i$; in this notation, these are also
written as
$$ D_i \ = \ \pa_i \ + \ u^a_{J,i} \ \pa_a^J \ . $$

\subsection{Equations and symmetries}

We can now pass to describe our general framework (see again
\cite{\symmref} for details). Consider a bundle $(M,\pi,B)$ with
fiber $\pi^{-1} (x) = U$; we will use local coordinates
$(x^1,...,x^q)$ on the base manifold $B$, and $(u^1,...,u^p)$ on
the fiber $U$; a function $u = f(x)$ corresponds to a section
$\ga_f$ of $(M,\pi,B)$. Associated to $(M,\pi,B)$ are the Jet
bundles $J^n M$; natural local coordinates on these are provided
by $(x,u)$ and partial derivatives (up to order $n$) of the $u$
with respect to the $x$. The Jet bundles are naturally endowed
with a {\it contact structure} $\mathcal{C}$; this is generated by
the {\it contact forms}, given in local coordinates by
$$ \om^a_J \ := \ \d u^a_J \ - \ u^a_{J,i} \, \d x^i \ . $$

A function $u = f(x)$ also identifies its partial derivatives of
any order; in the same way, a section $\ga_f$ of $(M,\pi,B)$ also
identifies a section $\ga_f^{(n)}$ of $(J^n M , \pi_n , B)$, i.e.
of $J^n M$ seen as a bundle over $B$. The section $\ga_f^{(n)} \in
\Sigma (J^n M)$ is said to be the \emph{prolongation} of $\ga_f \in
\Sigma (M)$.

A differential equation, or system thereof, $\De$ (of order $n$)
is a relation involving the independent variables $x$, the
dependent variables $u$, and their derivatives (up to order $n$);
thus it defines a \emph{solution manifold} $S_\De$ in $J^n M$. The
function $u = f(x)$ is a solution to $\De$ if and only if the
prolongation $\ga_f^{(n)}$ of $\ga_f$ lies entirely in, i.e. is a submanifold of, $S_\De$.

Let us now consider a vector field $X$ in $M$; this is written in
coordinates as \beq\label{eq:X} X \ = \ \xi^i (x,u) \, \pa_i \ + \
\phi^a (x,u) \, \pa_a \ . \eeq This generates a (local)
one-parameter group of diffeomorphisms in $M$; its action on
$\Sigma (M)$ is described by its \emph{evolutionary representative}
\beq\label{eq:Xv} X_v \ = \ (\phi^a \ - \ u^a_i \, \xi^i ) \ \pa_a
\ := \ Q^a (x,u,u_x) \ \pa_a \ . \eeq That is, the infinitesimal
action of $X$ maps the function $u = f(x)$ into the function $u =
\^f (x)$, with $\^f (x) = f(x) + \eps [ \phi^a (x,u) - u^a_i \xi^i
(x,u) ]$, where $u$ and $u_i$ should be computed on $u = f(x)$.

The action of $X$ on $M$ induces an action on $J^n M$; this is
described by the vector field \beq\label{eq:Xn} X^{(n)} \ = \
\xi^i (x,u) \, \pa_i \ + \ \psi^a_J (x,u,..., u^{(|J|)} ) \,
\pa_a^J \eeq where the coefficients $\psi^a_J$ depend on partial
derivatives $u^a_K$ of order up to $|J|$ and satisfy, with
$\psi^a_0 \equiv \phi^a$, the {\it prolongation formula}
\beq\label{eq:prol} \psi^a_{J,i} \ = \ D_i \psi^a_J \ - \
u^a_{J,k} \ D_i \xi^k \ . \eeq

The vector field $X$ is said to be a \emph{symmetry generator}, or (with
a slight abuse of language) simply a \emph{symmetry} for $\De$ if
it (locally, i.e. near to zero in the group parameter) maps
solutions into solutions; this is seen \cite{\symmref} to be
equivalent to the following

\medskip\noindent
{\bf Definition 1.} {\it The vector field $X$ in $M$ is a {\rm symmetry
generator} for the differential equation $\De$ of order $n$ if and
only if $X^{(n)} :  S_\De \to \T S_\De.$}

\medskip\noindent
{\bf Remark 1.} Note that if $Y = c X$ with $c$ a constant, then
$Y^{(n)} = c X^{(n)}$. Thus if $X$ is a symmetry of $\De$, so is
$Y = c X$ for any $c \in {\bf R}$.

On the other hand, if $Y$ is collinear but not
proportional to $X$, i.e. $Y = f(x,u) X$ with $f$ non constant,
then their prolongations do {\it not} satisfy the same relations,
and are in general not collinear. This is readily seen considering
first prolongations, which is enough to make our point. In fact,
for $X$ given by \eqref{eq:X}, we have $Y = \eta^i \pa_i + \b^a
\pa_a$, with $\eta^i = f \xi^i$ and $\b^a = f \phi^a$. The
prolongation formula \eqref{eq:prol} provides then $X^{(1)} = X +
\psi^a_i \pa_a^i$ and $Y^{(1)} = Y + \chi^a_i \pa_a^i$ with
$\psi^a_i = D_i \phi^a - u^a_k D_i \xi^k$ and
\begin{eqnarray*} \chi^a_i &=& D_i \b^a \ - \ u^a_k \ D_i \eta^k \ = \
D_i (f \, \phi^a) \ - \ u^a_k \ D_i (f \xi^k) \\
&=& f \ \( D_i \phi^a \ - \ u^a_k \, D_i \xi^k \) \ + \ (\phi^a \
- \ u^a_k \, \xi^k) \ (D_i f) \ . \end{eqnarray*} Thus, writing
$Q^a = (\phi^a - u^a_k \xi^k)$, we have
$$ Y^{(1)} \ = \ f \ X^{(1)} \ + \ Q^a \, (D_i f) \, \pa_a^i \ = \
f \ \[ X^{(1)} \ + \ f^{-1} (D_i f) \, Q^a \, \pa_a^i \] \ . $$
For higher order prolongations, we get a similar result; see the
Appendix A. \EOR

\subsection{Sets of vector fields}

Let us now consider a set of vector fields $X_\a$, written in
coordinates as \beq\label{eq:Xa} X_\a \ = \ \xi^i_\a (x,u) \,
\pa_i \ + \ \phi^a_\a (x,u) \, \pa_a  \eeq and with prolongations
\beq\label{eq:Xan} X_\a^{(n)} \ = \ \xi^i_\a (x,u) \, \pa_i \ + \
\psi^a_{\a;J} (x,u,..., u^{(|J|)} ) \, \pa_a^J \ . \eeq

We are interested in the case where the $X_\a$ are in {\it
involution} (in Frobenius sense), i.e. the set is closed under
commutation: \beq \label{eq.invo} [X_\a , X_\b ] \ = \ F_{\a
\b}^\ga (x,u) \ X_\ga \ , \eeq with $F_{\a \b}^\ga$ smooth
functions on $M$. If the $F_{\a \b}^\ga (x,u)$ are actually
constant (we will then write them as $c_{\a \b}^\ga$) we have a
\emph{Lie algebra} of vector fields.

It is well known \cite{\symmref} that if $X,Y$ are vector fields on $M$, and
$X^{(n)} , Y^{(n)}$ their prolongations, then \beq \[ X^{(n)} ,
Y^{(n)} \] \ = \ \( [X,Y] \)^{(n)} \ . \eeq
In a slightly different formulation,
$$ [X,Y] \ = \ Z \ \Rightarrow \ \ [X^{(n)} , Y^{(n)} ] \ = \ Z^{(n)} \ . $$

It follows easily from this that if the vector fields $X_\a$ span a Lie
algebra, then their prolongations span the same Lie algebra; in fact,
$$ \[ X_\a^{(n)} , X_\b^{(n)} \] \ = \ \( [X_\a,X_\b] \)^{(n)} \ = \
\( c_{\a \b}^\ga \ X_\ga \) ^{(n)} \ = \ c_{\a \b}^\ga \ X_\ga^{(n)} \ . $$
Note this depends on the $c_{\a \b}^\ga$ being constant.\footnote{We recall in passing that, as well known, the symmetry generators of a given $\De$ form a
Lie algebra under commutation. Note that this could be
infinite-dimensional as a Lie algebra (albeit spanning a
finite-dimensional distribution, as guaranteed by the finite
dimensionality of $M$ and of $J^n M$).}

\bigskip\noindent
{\bf Remark 2.} On the other hand, if the $X_\a$ are in involution
but do {\it not} span a Lie algebra (that is, at least some of the
$F_{\a \b}^\ga$ are not constant), then their prolongations do
{\it not} satisfy the same involution relations, and could very
well not be in involution. More precisely, we have that if the
$X_\a$ satisfy eq. \eqref{eq.invo}, then \beq \[ X_\a^{(n)} ,
X_\b^{(n)} \] \ = \ \( [X_\a,X_\b] \)^{(n)} \ + \ Z_{\a \b}^{(n)}
\ , \eeq where vector field $ Z_{\a \b}^{(n)}$ is nonzero; the
recursion relation obeyed by its coefficients will be determined
in the Appendix A.  \EOR

\medskip\noindent
{\bf Remark 3.} Let $X_\a$ be symmetries of $\De$. It is clear
that for any choice of smooth functions $f_\a : J^n M \to \R$, the
vector field $Z = f_\a X_\a^{(n)}$ is tangent to $S_\De$. On the
other hand, a generic vector field in $J^n M$ of this form, for
$f$ non constant and generic, will \emph{not}  be the prolongation
of \emph{any } vector field in $M$, and surely not of the vector
field $Z_0 = f_\a X_\a$. Note that -- as we will discuss later on
-- there will be special choices of non-constant $f_\a$ for which $Z$ is a
prolongation of some vector field $W \not= Z_0$ in $M$. \EOR

\section{An equivalent formulation of the symmetry condition}
\label{sec:equiv}

The discussion of the previous section suggests to shift focus in
our description of symmetries (keeping of course the same
meaning); that is, rather than focusing on a vector field $X$, or
set of vector fields $\mathcal{X} = \{ X_1 , ... , X_r \}$, we
will look at the {\bf distribution} $\D_X$ or $\D_\mathcal{X}$
their prolongation span in $\T J^n M$, also called the \emph{prolonged distribution}, or more precisely at the integral manifolds for these distributions \cite{God}.

In the case of a single vector field, $\D_X$ is one-dimensional
(note it may have singular points, corresponding to singular
points -- i.e. zeros -- of $X$); for a set $\X$, the dimension of
$\D_\X$ corresponds to the rank of $\X$ and again we can have
singular points where $\D_\X$ has a lower dimension.

\medskip\noindent
{\bf Definition 2.} {\it The vector field $X$ in $M$ is a {\rm symmetry
generator} for the differential equation $\De$ of order $n$ if
$S_\De \ss J^n M$ is an integral manifold for the distribution
$\D_X$ spanned by $X^{(n)}$ in $J^n M$.  The set $\X = \{ X_1 ,
... X_r \}$ of vector fields in $M$ is a {\rm set of symmetry
generators} for the differential equation $\De$ of order $n$ if
$S_\De \ss J^n M$ is an integral manifold for the distribution
$\D_\X$ spanned by $X_\a^{(n)}$ in $J^n M$.}
\bigskip

This definition is equivalent to the previous one, but shifts our
attention from the vector fields to the distribution they
generate.

It is quite obvious that if we drop the requirement to have
prolonged vector fields we have many different sets of generators
for $\D_X$ or $\D_\X$;
but in this case the symmetry is, in general, of little use for
the usual goals of symmetry analysis; that is, for reducing the
equation under study (in the case of ODEs), or at least for
determining its invariant solutions (for PDEs).

We thus wonder if there is some intermediate class: vector fields
in $J^n M$ which are not prolongations of vector fields in $M$,
but which can still be used for symmetry reduction. As mentioned
above, one should distinguish between ODEs and PDEs
\medskip

{\bf (1)} Let us first consider ODEs. Looking at the standard
symmetry reduction procedure for ODEs \cite{\symmref}, we realize
this is based on the so called {\it Invariant By Differentiation
Property} (IBDP). This means that once we know differential
invariants of order zero (call them $\eta$) and one (call them
$\zeta^{(1)}$) for a vector field, differential invariants of
order two can be computed simply by
$$ \zeta^{(2)} \ = \ \frac{D_x \zeta^{(1)}}{D_x \eta} \ ; $$
the procedure is then iterated to generate higher order
invariants. Note in this way we generate (for ODEs, i.e. for a
single independent variable $x$) a complete set of differential
invariants of any given order.

It should be remarked that differential invariants will be the
same for the vector field $Y = X^{(n)}$ and for $Z = f Y$, for any
smooth function $f$ which is nowhere vanishing, or at least which
vanishes only at singular points of $Y$. This shows immediately
that the IBDP property will hold true for any vector field of the
form $Z = f Y$, i.e. that (as it could be expected) symmetry
reduction is still possible if we consider a different generator
for the one-dimensional distribution generated by $Y = X^{(n)}$ in
$J^n M$.

It is then natural to ask if this vector field $Z = f Y$ can be
associated in some way to a vector field $W$ in $M$, through some
kind of deformed prolongation operation.

Similar considerations hold for sets of vector fields, i.e. for $Y_\a = X_\a^{(n)}$ and $Z_\a = F_\a^{\ \b} Y_\b$ with $F$ a smooth regular matrix function.
\medskip

{\bf (2)} As for determining invariant solutions to PDEs (by this
we mean solutions which are invariant under some subalgebra $\G_0$
of the full symmetry algebra $\G$ of the equation; one could
actually also consider more general settings
\cite{ClK,LeWi1,CGpart}, but here we will not consider these), this
is based on the restriction of the equation to the set of
functions which are invariant under the symmetry vector field. The
restriction of this equation corresponds, in geometrical terms, to
the intersection of the solution manifold $S_\De$ with the fixed
point set $\mathcal{F}_0$ for all the vector fields $Y_\a =
X_\a^{(n)}$ given by prolongations of the generators $\X = \{ X_\a
\}$ of the invariance subalgebra $\G_0$. From our point of view
$\mathcal{F}_0$ is the set of maximally singular points for the
distribution $\D_\X$. Again, it is clear that changing the system
of generators by passing to $Z_\a = F_\a^{\ \b} Y_\b$ for this
distribution does not change $\mathcal{F}_0$, and hence the
invariant solutions $S_\De \cap \mathcal{F}_0$.

In this context too it is natural to ask if these vector
field $Z_\a = F_\a^{\ \b} Y_\b$ can be associated in some way to
vector fields $W_\a$ in $M$, through some kind of deformed
prolongation operation.
\bigskip

The question raised in points (1) and (2) above appears to be new,
but it has actually been implicitly solved (working from a
different point of view and with a different focus) in the
literature dealing with {\it twisted prolongations} of vector
fields and {\it twisted prolongations}.

\medskip\noindent
{\bf Remark 4.} The field of twisted symmetries of differential
equations was started by the works of Muriel \& Romero with the so
called $\lambda$-symmetries \cite{MuRom1}, and by the geometrical
understanding of these by Pucci \& Saccomandi \cite{PuS}; the
scope of the theory, initially limited to scalar ODEs, was then
extended to systems of ODEs \cite{MuRom4} and to PDEs
\cite{CGMor,GMor}, and recently also to sets of vector fields
rather than single ones \cite{Sprol,SprolDS,CGWjlt1,CGWjlt2}. See
\cite{Gtwist1,Gtwist2} for reviews on twisted prolongations and
twisted symmetries. Roughly speaking, one consider a modification
(twisting) of the standard prolongation rule, depending on an
auxiliary object\footnote{This may be a scalar function $\la
(x,u,u_x)$ in $\la$-symmetries, a set of $q$ $(p \times p)$ matrix
functions $\La_i (x,u^{(1)}$ defining the semi-basic one-form $\mu
= \La_i \d x^i$ satisfying the horizontal Maurer-Cartan equation
as in $\mu$-symmetries, or a matrix-valued function $\s^{\a \b}
(x,u^{(1)} ) $ (here the indices refer to a generating set of
vector fields \cite{Sprol,SprolDS}) as in $\s$-symmetries.}, but
such that the vector fields whose such twisted prolongations is tangent to $S_\De$ can still be used in the same way as standard symmetries for
reducing the differential equations under study (in the case of
ODEs or systems thereof) or to obtain invariant solutions (in the
case of PDEs or systems thereof). The fact that twisted symmetries
are as useful as standard ones for these tasks is due to certain
algebraic facts (for ODEs and reduction, to satisfying the
Invariant By Differentiation Property (IBDP) \cite{MuRom3}; for
PDEs, to the fact twisted prolongations coincide with standard
ones on the set of invariant functions \cite{CGMor,GMor}); we
trust the present work sheds light on the geometrical reason
behind these. \EOR

\medskip\noindent
{\bf Remark 5.} It should be stressed that ''twisted symmetry'' is in a a way a misnamer: a twisted symmetry does in general \emph{not} map solutions into solutions, nd hence is not, properly speaking, a symmetry \cite{\symmref} (but for PDEs, see point (2) above, when they can be applied to reduce the equation they are at least conditional \cite{LeWi1}, or partial \cite{CGpart}, symmetries). Moreover while standard symmetries can be computed algorithmically, this is not the case for twisted symmetries: their determination relies in general on guesswork or physical considerations, albeit in some cases the analysis of the system under study can provide insights for the structure of twisted symmetries \cite{SprolDS}. \EOR

\section{Distributions and twisted symmetries}
\label{sec:distrib}

As mentioned above, we will use some results from investigations
on twisted symmetries; we will quote these from the review paper
\cite{Gtwist2}. It turns out the results are more simply
stated in terms of evolutionary representatives of vector fields,
see \eqref{eq:Xv} above. We will find that changing the set of generators in the prolonged distributions associated to a vector field or a set of vector fields leads quite naturally to consider twisted prolongations and symmetries.

\subsection{Single vector fields}

We start by considering a single vector field $X$, which we write
in the form \eqref{eq:X}. In this context, we consider its
$\mu$-prolongation, based on a semi-basic one form $\mu$ on $(J^1
M,\pi_1,B)$, which we write as
$$ \mu \ = \ \La_i (x,u^{(1)} ) \ \d x^i \ ; $$ the $\La_i$ are
$(p \times p)$ matrices (recall $p$ is the dimension of the fiber
$\pi^{-1} (x)$ in $M$), satisfying the compatibility conditions
\beq\label{eq:hMC} D_i \, \La_j \ - \ D_j \, \La_i \ + \ [\La_i ,
\La_j ] \ = \ 0 \ . \eeq The latter is just the horizontal
Maurer-Cartan equation \cite{CCL,Sha}. (Dealing with the
horizontal version of this is rather natural in view of the
presence of the contact structure in $J^n M$.) By introducing the
operators $$ \nabla_i \ := \ D_i \ + \ \La_i \ , $$ defined more precisely as $ (\nabla_i)^a_{\ b} = (\de^a_{\ b} ) D_i + (\La_i)^a_{\ b}$, this is also
reformulated as the zero-curvature condition \cite{CFX,FT,Mar}
$$ [\nabla_i \, , \, \nabla_j ] \ = \ 0 \ . $$

The $\mu$-prolongation of the vector field $X$ is defined as the
vector field \eqref{eq:Xn} with coefficients $\psi^a_J$ satisfying
the {\it $\mu$-prolongation formula} \beq \label{eq:muprol}
\psi^a_{J,i} \ = \ D_i \psi^a_J \ - \ u^a_{J,k} \, D_i \xi^k \ + \
(\La_i)^a_{\ b} \ \( \psi^b_J \ - \ u^b_k \, \xi^k \) \ ; \eeq
this is also rewritten in terms of the $\nabla_i$ defined above as
$$ \psi^a_{J,i} \ = \
(\nabla_i)^a_{\ b} \, \psi^b_J \ - \ u^b_{J,k} \, (\nabla_i)^a_{\
b} \, \xi^k \ . $$ In order to stress the application of the
modified prolongation operation, we will denote the $n$-th
$\mu$-prolongation of $X$ as $X^{(n)}_\mu$.

\medskip\noindent
{\bf Remark 6.}
Note that for ODEs we have only one matrix $\La$; a special case occurs, as discussed below, when $\La$ is a multiple of the identity via a smooth function $\la : J^1 M \to R$, i.e. for $\La = \la I$. In this case we are reduced to the setting of $\la$-prolongations and symmetries introduced by Muriel and Romero for scalar equations \cite{MuRom1,MuRom2,MuRom3} and for systems\cite{MuRom4}. Thus $\la$-prolongations and symmetries can be seen, for the sake of our present discussion, as special cases of $\mu$-prolongations and symmetries. (The same would apply for $\s$-prolongations and symmetries to be considered below; in that case we obtain $\la$ ones when the set reduces to a single vector field and hence the matrix $\s$ to a scalar function $\la$.) However, this special case has some special -- and quite convenient -- features, see below. \EOR

\medskip\noindent
{\bf Lemma 1.} {\it Consider the evolutionary vector fields $X$
and $\wt{X}$ on $M$ given by  $X =Q^a \pa_a$ and $\wt{X} = (A^a_{\
b} Q^b ) \pa_a$, with $A : M \to \mathtt{Mat} (\R,q)$ a nowhere
zero smooth ($C^\infty$) matrix function on $M$. Consider moreover
the standard prolongation of $\wt{X}$ and the $\mu$-prolongation
of $X$ with $\mu = A^{-1} (D A)$; i.e. the vector fields $
X^{(n)}_\mu = Q^a_J \pa_a^J$ and $\wt{X}^{(n)} = \wt{Q}^a_J
\pa_a^J$ in $J^n M$. Then \beq\label{eq:L6}  A^a_{\ b} \ Q^b_J \ =
\ \wt{Q}^a_J \ . \eeq}

\medskip\noindent
{\bf Proof.} First of all we note that by definition the $Q^a_J$
obey the $\mu$-prolongation formula, and the $\wt{Q}^a_J$ the
standard one, so that
$$ Q^a_{J,i} \ = \ D_i \, Q^a_J \ + \ (\La_i)^a_{\ b} \, Q^b_J \ , \ \
\wt{Q}^a_{J,i} \ = \ D_i \, \wt{Q}^q_J \ . $$ Let us write, for
all $J$,
$Q^a_J = L^a_{\ b} P^b_J$,
with $L$ a $C^\infty$ and nowhere singular matrix function on $M$.
Then the $\mu$-pro\-lon\-ga\-tion formula requires
\begin{eqnarray*}
Q^a_{J,i} &=& D_i (Q^a_J ) \ + \ (\La_i)^a_{\ b} \ Q^b_J \ = \
D_i (L^a_{\ b} \, P^b_J ) \ + \ (\La_i)^a_{\ b} \ L^b_{\ c} \, P^c_J  \\
&=& D_i (L^a_{\ b}) \, P^b_J \ + \ L^a_{\ b} \, D_i (P^b_J ) \ + \
(\La_i \, L)^a_{\ b} \, P^b_J  \ . \end{eqnarray*} On the other
hand, we know that $ Q^a_{J,i} = L^a_{\ b} P^b_{J,i}$; comparing
these two formulas, we get
$$ P^a_{J,i} \ = \ D_i \, P^a_J \ + \ [L^{-1} \, D_i (L )]^a_{\ b} \ P^b_J \ + \
(L^{-1} \, \La_i \, L )^a_{\ b} \, P^b_J \ . $$ We conclude that
the $P^a_J$ satisfy the standard prolongation formula -- and thus
can be identified with the $\wt{Q}^a_J$ -- provided $L$ satisfies
$ L^{-1} (D_i L) + L^{-1} \La_i L = 0$; equivalently, provided
$ (D_i L) L^{-1} = - \La_i$. Writing this relation in terms of $A = L^{-1}$, we have \beq
\label{eq:Lai} \La_i \ = \ A^{-1} \ (D_i A) \ . \eeq As $D A =
(D_i A) \d x^i$, the proof is completed. \EOP

\bigskip\noindent
{\bf Remark 7.} We can summarize the relations described by this
Lemma in the form of a commutative diagram: \beq \label{diag:mu}
\matrix{ X & \mapright{A} & \wt{X} \cr
\mapdown{\mu-prol} & & \mapdown{prol} \cr
\ X^{(n)}_\mu & \mapright{A} & \ \wt{X}^{(n)}_0 \cr} \eeq

The matrices $\La_i$ defining $\mu$ and $A$ are related by
\eqref{eq:Lai}. The $C^\infty$ smoothness of $A$ entails
$C^\infty$ smoothness of the $\La_i$. In the case $\La_i = \la_i
I$, the equivalence is through a simple rescaling of the vector
fields. The case of $\la$-symmetries is a special case of that of
$\mu$-symmetries. \EOR

\medskip\noindent
{\bf Remark 8.} Needless to say, the diagram can also be read the other way round. That is, considering the map $L = A^{-1}$, we state that acting with the map $L$ on $\wt{X}_0^{(n)}$ (and on $\wt{X}$) we obtain a vector field $L \circ \wt{X}_0^{(n)}$ which is the $\la$-prolongation of the vector field $X = L \circ \wt{X}$. \EOR

\bigskip\noindent
{\bf Remark 9.} Here the matrix $A$ acts on the vector indices
corresponding to the variables of the space tangent to the fiber
$U = \pi^{-1} (x)$, i.e. -- in terms of local coordinates -- to
the dependent variables $u^a \in U$ and their derivatives $u^a_J$,
belonging (for each given $J$) to a vector space $U_J$ isomorphic
to $U$. \EOR

\bigskip\noindent
{\bf Remark 10.} The horizontal Maurer-Cartan equation
\eqref{eq:hMC} is automatically satisfied for $\La_i$ given by
\eqref{eq:Lai}. \EOR

\bigskip\noindent
{\bf Remark 11.} Lemma 1 above deals with evolutionary
representatives, which are generalized vector fields in $M$;
however while $X = Q^a \pa_a$ (recall here $Q^a = \phi^a - u^a_k
\xi^k$) is by construction the evolutionary representative of the
Lie-point vector field $X_0 = \xi^i \pa_i + \phi^a \pa_a$, it
is not at all clear that $\wt{X} = (A^a_{\ b} Q^b) \pa_a$ is the
evolutionary representative of a Lie-point vector field
$\wt{X}_0$ in $M$. The evolutionary representatives
\eqref{eq:Xv} can be characterized, among first order generalized
vector fields on $M$ \cite{Olv1}, as those satisfying
$$ \pa Q^a / \pa u^b_k \ = \ - \, \de^a_b \ \xi^k \ , \ \ Q^a
\ + \ \( \pa Q^a / \pa u^b_k \) \ u^b_k \ = \ \phi^a \ ; $$
note in particular that $(\pa Q^a / \pa u^a_k)$ (no sum on $a$) is
independent of $a$. By applying these requests on $\^X = \^Q^a
\pa_a$, $\^Q^a = A^a_{\ b} Q^b$, we obtain that while we always
have $\^Q^a + (\pa \^Q^a / \pa u^b_k ) u^b_k = \vartheta^a =
A^a_{\ b} \phi^b$, the first requirement is satisfied only for $A
= \la I$, or for all the $\xi^k$ identically vanishing. The first
case amounts to a rescaling of vector fields, while the second
applies when the considered symmetries do not act on independent
variables. (We note in passing that examples of application of
$\mu$-symmetries seemingly always refer to either one of these cases.) \EOR
\bigskip

We note now that the vector fields $X_\mu^{(n)}$ and $\wt{X}_0^{(n)}$ are in general \emph{not} collinear, and hence do \emph{not} span the same distribution. Thus, albeit a $\mu$-prolongation is associated to a standardly prolonged (local or nonlocal, see Section \ref{sec:nonlocal}) vector field, these are in general not in the relation of interest here, see Section \ref{sec:equiv}.

On the other hand, \emph{if} $A = \la I$ (we will then write $X_\la^{(n)}$ for the twisted prolongations), then $\wt{X}$ is collinear to $X$, and $\wt{X}_0^{(n)}$ is collinear to $X_\la^{(n)}$: we are then in the case where the prolonged distributions $\D_{\wt{X}}^{0}$ and $\D_X^\la$ do coincide. Recalling that in this case we deal actually with $\la$-prolongations and $\la$-symmetries, we can conclude that Lemma 1 above has a simple but relevant corollary:

\medskip\noindent
{\bf Corollary 1.} {\it Let $X$ be a $\la$-symmetry for the ordinary differential equation (or system thereof) $\De$. Then locally there is a (possibly nonlocal) vector field $\wt{X} = A \circ X$ which is a standard symmetry for the same equation, with $A = \a I$ and the relation between $\a$ and $\la$ is given by
$\la = \a^{-1} D_x \a$.

Conversely, if $\wt{X}$ is a standard symmetry for $\De$, applying a transformation $A^{-1} = \a^{-1} I$ we obtain a vector field $X$ which is a $\la$-symmetry for $\De$, with again the same relation between $\a$ and $\la$.} \bigskip

It should be noted that twisting the prolongation operation has
little effect on the set of invariant functions, and thus it is
not unexpected that it also preserves the possibility of reduction
of PDEs. In fact by this one usually means the search for
invariant solutions, i.e. of solutions obtained by restricting the
PDE to the set of $X$-invariant sections (i.e. $X$-invariant
functions).

\medskip\noindent
{\bf Lemma 2.} {\it Consider the evolutionary vector fields $X =
Q^a \pa_a$ and $\wt{X} = (A^a_{\ b} Q^b ) \pa_a$ on $M$, with $A$
as in Lemma 1. Let $\mathcal{F}_X$ denote the set of $X$-invariant
sections in $\Sigma (M)$ (i.e. of $X$-invariant functions $f : B
\to U$). Then $X^{(n)}_\mu$, $\wt{X}^{(n)}$ and $X^{(n)}_0$ coincide on
$\mathcal{F}_X$.}

\medskip\noindent
{\bf Proof.} First of all we note that invariant sections $\ga$
are characterized precisely by the fact that all the $Q^a$ vanish
on them. Thus $X^{2)}_0$ is surely null on $\mathcal{F}_X$.

It was shown in \cite{GMor} (see Theorem 3 in there) that writing
the standard and $\mu$-prolonged vector fields of the same
evolutionary vector field $X = \phi^a \pa_a$ in the form
$$ X^{(n)} \ = \ \psi^a_J \, \pa_J \ , \ \ X_\mu^{(n)} \ = \
\[ \psi^a_J \ + \ F^a_J \] \, \pa_a^J \ , $$
the difference term $F^a_J$ satisfies the recursion relation
$$ F^a_{J,i} = \( \de^a_b \, D_i \ + \ (\La_i)^a_{\ b} \) \, F^b_J \ + \
(\La_i) ^a_{\ b} \ D_J \, Q^b \ . $$ This is started by $F^a_0 =
0$. It is thus clear that $F^a_J$ vanishes on $\mathcal{F}_X$ for
all $J$, hence $X_\mu^{(n)} = X_0^{(n)}$ on $\mathcal{F}_X$.

To conclude the proof, it suffices to note that the
characteristics $Q$ of $X$ and $\wt{Q}$ of $\wt{X}$ are related by
a linear (point-dependent) invertible transformation $\wt{Q}^a =
A^a_{\ b} Q^b$, hence they vanish on the same set. \EOP
\bigskip

Recalling point (2) in the discussion of Section \ref{sec:equiv}, we immediately conclude that if an equation (or system thereof) $\De$ admits a vector field $X$ as a $\mu$-symmetry, this can be used exactly as standard symmetries in the determination of special (invariant) solutions.

Actually Lemma 2 allows to be more specific and reduce the situation to the familiar one (albeit in practice it may be more convenient to avoid this step). In fact, we have again a simple but useful corollary.

\medskip\noindent
{\bf Corollary 2.} {\it Let the equation (or system thereof) $\De$ admit $X$ as a $\mu$ symmetry, with $\mu = \La_i \d x^i$; and let $X$-invariant solutions exist for $\De$. Then there is a vector field $\wt{X} = A \circ X$ which is a conditional symmetry for $\De$, and the $X$-invariant solutions are also $\wt{X}$-invariant. The relation between $\mu$ and $A$ is given by \eqref{eq:L6}.}
\bigskip

This Corollary allows to reduce the search for special solutions associated to $\mu$-symmetries to the familiar case of special solution associated to standard conditional symmetries \cite{LeWi1}.

\subsection{Sets of vector fields}

When we consider a set $\X$ of vector fields $X_\a$, we can of
course apply to each one of them the considerations presented in
the previous subsection. However, in this case the generators of
the distribution $\D_\X$ can also be changed by ``mixing'' the
prolongations of the different vector fields. This is the case we
are considering now, restricting to the case of ODEs (one independent variable, denoted as $x$).

\medskip\noindent
{\bf Lemma 3.} {\it Let $\mathcal{X} = \{ X_1 , ... , X_r \}$ be a
set of vector fields on $M$; and let the vector fields
$\mathcal{Y} = \{ Y_1 , ... , Y_r \}$ on $J^n M$ be their
$\s$-prolongation. Consider also the set $\mathcal{W} = \{ W_1 ,
... , W_r \}$ of vector fields on $M$ given by $W_\a = A_\a^{\ \b}
X_\b$, with $A$ a nowhere singular matrix function on $M$; and let
the vector fields $\mathcal{Z} = \{ Z_1 , ... , Z_r \}$ on $J^n M$
be their standard prolongation. Then, provided $A$ and $\s$ are
related by \beq\label{eq:L11} \s \ = \ A^{-1} \ D_x A \ , \eeq we
also have $Z_\a = A_\a^{\ \b} Y_\b$.}

\medskip\noindent
{\bf Proof.} It will suffice to consider first prolongations. In
coordinates and with the usual shorthand notation,
\begin{eqnarray*} X_\a &=& \xi_\a \, \pa_x \ + \ \phi^a_\a \,
\pa_a \ , \\ Y_\a &=& X_\a \ + \ \( (D_x \phi^a_\a \, - \, u^a_x
\, D_x \xi_\a) \ + \ \s_\a^{\ \b} \, (\phi^a_\b \, - \, u^a_x \,
\xi_\b ) \) \, \pa_a^1 \ ; \\
W_\a &=& \chi_\a \, \pa_x \ + \ \eta^a_\a \, \pa_a \ , \\ Z_\a &=&
W_\a \ + \ (D_x \eta^a_\a \, - \, u^a_x \, D_x \chi_\a) \, \pa_a^1
\ . \end{eqnarray*}
If now we require $W = A X$, i.e.
$$ \chi_\a \ = \ A_\a^{\ \b} \, \xi_\b \ ; \ \
\eta^a_\a \ = \ A_\a^{\ \b} \, \phi^a_\b \ , $$ we immediately get
\begin{eqnarray*}
D_x \chi_\a &=& (D_x A_\a^{\ \b}) \, \xi_\b \ + \ A_\a^{\ \b} \,
(D_x \xi_\b) \ , \\
D_x \eta^a_\a &=& (D_x A_\a^{\ \b}) \, \phi^a_\b \ + \ A_\a^{\ \b}
\, (D_x \phi^a_\b) \ . \end{eqnarray*} Inserting these in the
expression for $Z$, we get \begin{eqnarray*} Z &=& A_\a^{\ \b} \
\[ \xi_\b \, \pa_x \ + \ \phi^a_\b \, \pa_a \right. \\
& & \left. \ \ \ + \ \( (D_x \phi^a_\b - u^a_x D_x \xi_\b) \ + \
(A^{\-1} \, D_x A)_\a^{\ \b}
\, (\phi^a_\b - u^a_x \xi_\b ) \) \] \, \pa_a^1 \\
&=& A_\a^{\ \b} \, Y_\b \ + \ \[ (D_x A)_\a^{\ \b} \ - \ (A
\s)_\a^{\ \b} \] \, (\phi^a_\b - u^a_x \xi_\b ) \, \pa_a^1 \ .
\end{eqnarray*} Thus, provided $A$ and $\s$ satisfy $D_x A = A
\s$, and hence (recalling $A$ is nowhere singular) satisfy
\eqref{eq:L11}, we have that $W = A X$ leads to $Y = A Z$. \EOP
\bigskip

This Lemma shows that the distributions in $\T J^n M$ generated by $\mathcal{Y}$ and by $\mathcal{Z}$ do coincide. We have then immediately an analogue of Corollary 1 for the case of systems of vector fields.

\medskip\noindent
{\bf Corollary 3.} {\it Let the set $\{ X_i \}$ of vector fields be a $\s$-symmetry for the differential equation (or system thereof) $\De$. Then locally there is a set $\{ W_i \}$, with $W_i = A_i^{\ j} X_j$ of vector fields which is a set of standard symmetries for the same equation; the relation between $A$ and $\s$ is given by \eqref{eq:L11}.

Conversely, if $\{W_i \}$ is an involution system of standard symmetries for $\De$, applying a transformation $A^{-1}$ we obtain a set $\{ X_i \}$ of vector fields which is a $\s$-symmetry for $\De$, with again the same relation between $A$ and $\s$.} \bigskip

\medskip\noindent
{\bf Remark 12.} The relations between the vector fields $X_i$ and
$W_i$, and their (respectively, $\s$ and standard) prolongations,
as given by Lemma 3, can be summarized in the form of a
commutative diagram: \beq \label{diag:sigma} \matrix{ \{ X_i \} &
\mapright{A} & \{ W_i \} \cr \mapdown{\s-prol} & & \mapdown{prol}
\cr \{ Y_i \} & \mapright{A} & \{ Z_i \} \cr} \eeq The relation
between $A$ and $\s$ is given by \eqref{eq:L11}. \EOR

\medskip\noindent
{\bf Remark 13.} As in the case of $\mu$-prolongations, the diagram can also be read the other way round: considering the map $L = A^{-1}$ we can say that to any involution set $\{ W_i \}$ of standard symmetries of an equation $\De$ is associated a set $\{ X_i \}$, with $X_i = L_i^{\ j} W_j$, which is a $\s$-symmetry of the same equation $\De$, the relation between $A = L^{-1}$ and $\s$ being given by \eqref{eq:L11}. \EOR

\medskip\noindent
{\bf Remark 14.} Lemma 3 stipulates that the relation between $A$
and $\s$ is given by \eqref{eq:L11}; note that if we look at this
as an equation for $A$ with a given $\s$, the solution is in
general not unique (an explicit example is provided in
\cite{Sprol}). Note also that the sets $\{Y_i \}$ and $\{ Z_i \}$
considered in Lemma 3 will in general have different involution
properties. A detailed discussion of this point is given in
\cite{Sprol,SprolDS}. \EOR

\medskip\noindent
{\bf Remark 15.} As mentioned above, if we consider a single vector field, the matrix $\s$ reduces to a scalar function, and the notion of $\s$-prolongations and symmetries reduce to that of $\la$-prolongations and symmetries. \EOR

\medskip\noindent
{\bf Remark 16.} The notion of $\s$-symmetry can be put in contact with the ''side conditions'' approach by Olver and Rosenau \cite{OR1,OR2}; on the other hand, the latter was used by Broadbridge, Chanu and Miller \cite{BCM} to study non-regular separation of variables. It is thus natural to wonder if $\s$-symmetries could be used to study the latter problem in general. I thank prof. Miller for pointing out the relation of $\s$-symmetries to his work. \EOR

\section{Local versus nonlocal equivalence}
\label{sec:nonlocal}

Our Lemma 1 and Lemma 3 above state that given a map $A$ applied
on vector fields and their prolongations, we obtain new vector
fields and, in general, their twisted prolongations. This relation
is summarized in the diagrams \eqref{diag:mu} and
\eqref{diag:sigma}, see Remarks 7 and 12; the relation between $A$
and $\mu$ or, respectively, $\s$ is given by \eqref{eq:Lai} and
\eqref{eq:L11}.

It should be noted, however, that these Lemmas consider the
situation where one starts from a given $A$, obtaining the
corresponding $\La_i$ or, respectively, $\s$. On the other hand,
equations \eqref{eq:Lai} and \eqref{eq:L11} suggest one could
consider the converse problem; that is, given a set of matrices
$\La_i$, or a matrix $\s$, wonder if there is an associated map
defined by $A$. The purpose of this section is indeed to briefly
discuss this point, i.e. the relations between the ``direct'' and
the ``inverse'' problem.

Note that while for a given function $A = A(x,u)$,
eq.\eqref{eq:Lai} obviously defines smooth local functions $\La_i
= \La_i (x,u,u_x)$, if we look for the $A$ which satisfy
\eqref{eq:Lai} for given $\La_i$, it is much less obvious that a
solution exists, and even less that $A$ will be a local function
of $(x,u)$. The first fact is guaranteed by the (horizontal)
Maurer-Cartan equation \eqref{eq:hMC}, but the second is in
general not true.

In fact, the solution to \eqref{eq:Lai} is written formally as
\beq\label{eq:Aformal} A \ = \ \exp \[ \int \La_i \cdot \d x^i \]
\ ; \eeq unless the $\La_i$ are themselves obtained as total
derivatives of some ``pseudo-potential'', $\La_i = D_i \Phi$, this
will \emph{not} be a local function of the $x$ and $u$.

This point will be made clearer by an explicit example. For
two independent variables $(x,y)$ and one dependent variable $u$,
consider
$$ \La_x \ = \ \pmatrix{0 & u_x \cr 0 & 0 \cr} \ ; $$ one can check
that the most general form of $\La_y$ satisfying the horizontal
Maurer-Cartan equation \eqref{eq:hMC} is
$$ \La_y \ = \ \pmatrix{f_1 (x,y) - u f_2 (x,y) &
u_y - u^2 f_2 (x,y) + u [f_1 (x,y) - f_3 (x,y)] + f_4 (x,y) \cr
f_2 (x,y) & f_3 (x,y) + u f_2 (x,y) \cr} , $$
where the $f_i$ are arbitrary smooth functions of their argument.
If we choose e.g.
$$ \La_x \ = \ \pmatrix{ 0 & u_x \cr 0 & 0 \cr} \ , \ \
\La_y \ = \ \pmatrix{ 0 & u_y + h'(y) \cr 0 & 0 \cr} \ ; $$
these satisfy \eqref{eq:hMC}, and the $A$ identified by \eqref{eq:Lai} is simply
$$ A \ = \ \pmatrix{c_1 [u + h(y) ] & c_2 + c_3 [u + h(y) ] \cr c_1 & c_3 \cr} \ . $$
On the other hand, if we require that all the functions $f_i$ are
nowhere zero, e.g. set all them to be nonzero constants ($f_i x,y)
= k_i \not=0$), we get
$$ \Lambda_y \ = \ \pmatrix{k_1 - k_2 u & - k_2 u - k_2 u^2 + u_y + k_4 \cr k_2 & k_1 + k_3 + k_2 u \cr} \ ; $$
it is then easy to satisfy \eqref{eq:Lai} for $\La_x$ with a local
function $A=A(x,y,u)$; this is for any $A$ of the form
$$ A \ = \ \pmatrix{ \a_1 (y) + u \a_2 (y) & \a_3 (y) + u \a_4 (y) \cr  \a_2 (y) &  \a_4 (y)  \cr} \ . $$
But when we try to satisfy $D_y A = \La_y A$ for $A$ in this
class, the only possibility is given by $A=0$; that is, there is
no solution to \eqref{eq:Lai} with local functions.

\medskip\noindent
{\bf Remark 17.} It should be noted that in a previous paper of
mine and coauthors \cite{CGMor}, some result by Marvan \cite{Mar}
was incorrectly translated from the original mathematical language
of coverings and bi-complexes to the language used in that paper.
As a result, our paper gave a proposition (Proposition 2 in there)
attributed to Marvan which is not equivalent to what was proved by
Marvan in \cite{Mar} and which appears to be incorrect, as shown
above. \EOR

\medskip\noindent
{\bf Remark 18.} Note also that even in the ODE case, when only
one $x$ and one $\La$ are present and hence no Maurer-Cartan
condition is applied, the $A$ can very well be a non-local
function: this is e.g. definitely the case when $\La = \La (u)$.
This phenomenon is well known, and the equivalence of
$\la$-symmetries with non-local symmetries has been studied in the
literature \cite{CFM1,CFM2,CFM3,MuRom5,MuRom8,NuL}. \EOR

\medskip\noindent
{\bf Remark 19.} Note in this respect that in \cite{GJGP} and
\cite{GJPAgau} one considered gauge transformations that were
obtained by evaluating a certain $G$-valued function (with $G$ the
gauge group) on a section of the gauge bundle over $M$; this
amounted to a gauge fixing. Thus in this case one reduces to
evaluation of functions or integrals over a specific section. \EOR


\section{Examples}
\label{sec:examples}

We present some simple examples illustrating our discussion in concrete cases.

\bigskip\noindent
{\bf Example 1 ($\la$-symmetries).} In their seminal paper \cite{MuRom1}
Muriel \& Romero, following the discussion by Olver in his book \cite{Olv1},
considered equations of the form \beq \label{eq:MRexa} u_{xx} \ =
\ D_x \, F(x,u) \ . \eeq These can obviously be integrated by
quadratures, but may lack symmetries; this is e.g. the case for
\beq \label{eq:MREspec} F(x,u) \ = \ (x + x^2) \ e^u \ . \eeq

It was proven by Muriel \& Romero (see Theorem 4.1 in
\cite{MuRom1}) that any equation of the form \eqref{eq:MRexa}
admits the vector field $X = \pa / \pa u$ as a $\la$-symmetry with
the choice $\la = F_u (x,u)$; this allows to integrate the
equation via ($\la$) symmetry reduction. Thus in particular with
the choice \eqref{eq:MREspec}, i.e. for the equation
$$ u_{xx} \ = \ [1 + 2 x + u_x (x + x^2)] \ e^u \ , $$
we have $\pa/\pa u$ as a $\la$-symmetry with
$$ \la \ = \ (x + x^2) \ e^u \ . $$

The $\la$-prolonged vector field (acting in $J^2 M$, as we
consider a second order equation) turns out to be
$$ Y \ = \ \frac{\pa}{\pa u} \ + \ [ (x + x^2) \, e^u ] \, \frac{\pa}{\pa u_x}
\ + \ \[ \(e^u (x^2 +x )^2 +2 x + x u_x (x+1)+1\) \ e^u \] \,
\frac{\pa}{\pa u_{xx} } \ . $$

According to our Corollary 1, this is equivalent to a standardly
prolonged vector field $Z$, which is the prolongation of a
Lie-point vector field $W$; these are related to $Y$ and $X$ by a
function $\a (x,u)$ via $W = \a X$, $Z = \a Y$, and $\a$ satisfies
$\la = \a^{-1} (D_x \a)$.

This produces a non-local function $\a$, i.e.
$$ \a \ = \ \exp \[ \int (x + x^2) e^{u(x)} \ d x \] \ . $$

The vector field $W$ will then be simply $W = \a \pa_u$; its
standard prolongation turns out to be
$$ Z \ = \ \a \, \pa_u \ + \ (D_x \a) \, \pa_{u_x} \ + \ (D_x^2 \a
) \, \pa_{u_{xx}} \ ; $$ using $D_x \a = \a \la$ we get
$$ Z \ = \ \a \, \pa_u \ + \ \a \, \la \, \pa_{u_x} \ + \ \a \, (\la^2 + D_x
\la) \, \pa_{u_{xx}} \ = \ \a \, Y \ , $$ which confirms our
general result. In this case, due to the factor $\a$, $Z$ is of
course a non-local vector field (of exponential type) \cite{Olv1}.

\bigskip\noindent
{\bf Example 2 ($\mu$-symmetries).} Let us now consider an example of
$\mu$-prolongation for ODEs, hence with a single matrix $\La$,
keeping to the two-dimensional and second order case for the sake
of simplicity; we denote the independent variable by $x$, the
dependent ones by $(u,v)$.

We choose $\La$ as follows; the associated $A$ is then given by
\eqref{eq:L6}:
\beq\label{eq:EX2} \La \ = \ \pmatrix{0 & u_x \cr 0 & 0 \cr} \ ; \ \ \
A \ = \ \pmatrix{1 & u \cr 0 & 1 \cr} \ . \eeq Let us now consider
the vector field
$$ X \ = \ \pa / \pa v \ ; $$ its second order $\mu$-prolongation
turns out to be
$$ Y \ = \ X_\mu^{(2)} \ = \ \frac{\pa}{\pa v} \ + \ u_x \, \frac{\pa}{\pa u_x}
\ + \ u_{xx} \, \frac{\pa}{\pa u_{xx}} \ . $$

Differential invariants of order up to two for this vector field
are (note that, as mentioned in our general discussion, the IBDP
does not hold):
$$ x , \ u ; \ u_x \, e^{- v} , \ v_x ; \ u_{xx} \, e^{-v} ,  \
v_{xx} \ . $$

Thus $X$ is a $\mu$-symmetry for e.g. all the equations in the
class
\begin{eqnarray*}
u_{xx} &=& f_{11} (x,u) \, u_x \ + \ f_{12} (x,u) \, e^v \\
v_{xx} &=& f_{21} (x,u) \, v_x \ + \ f_{22} (x,u) \ ,
\end{eqnarray*}
with $f_{ij} (x,u)$ arbitrary smooth functions.

According to our Lemma 1, the vector field $Z = A \circ
X_\mu^{(2)}$ should be the standard prolongation of the vector
field $W = A \circ X$. This is indeed the case, as $W$ and $Z$ are
given respectively by
$$ W \ = \ u \, \frac{\pa}{\pa u} \ + \ \frac{\pa}{\pa v} \ ; \ \
\ \ Z \ = \ W \ + \ u_x \, \frac{\pa}{\pa u_x} \ + \ u_{xx} \,
\frac{\pa}{\pa u_xx} \ . $$ We note in passing that the
differential invariants of order up to two for $Z$ are
$$ x , \ u \, e^{- v} ; \ u_x/u , \ v_x ; \ u_{xx}/u , \ v_{xx} \
; $$ thus the IBDP holds for $Z$. Note that the differential
invariants for the vector fields $Y$ and $Z$ are not the same.

\bigskip\noindent
{\bf Example 3 ($\s$-symmetries).} In order to provide a simple example of $\s$-symmetries, again with one independent variable $x$ and two dependent variables $u,v$, we consider the scaling and rotation vector fields
given by
$$ X_1 \ = \ u \, \pa_u \ + \ v \, \pa_v \ , \ \ \ \ X_2 \ = \ v \, \pa_u \ - \ u \, \pa_v \ ; $$ and the matrix (which of course acts actually only on the first vector field)
$$ \s \ = \ \pmatrix{0 & u_x \cr 0 & 0 \cr} \ . $$
The second order $\s$-prolongation of the set $\X = \{ X_1 , X_2 \}$ is provided by $\Y = \{Y_1 , Y_2 \}$ with
\begin{eqnarray*}
Y_1 \ = \ (X_1)_\s^{(2)} &=& u \frac{\pa}{\pa u} +  v \frac{\pa}{\pa v} + u_x (1+ v) \frac{\pa}{\pa u_x} + (v_x - u u_x) \frac{\pa}{\pa v_x} \\
 & & \ +  [u_{xx} (1 + v) + 2 u_x v_x \frac{\pa}{\pa u_{xx}} + (v_{xx} - u u_{xx} - 2 u_x^2)  \frac{\pa}{\pa v_{xx}} \ , \\
Y_2 \ = \ (X_2)_\s^{(2)} &=& v \frac{\pa}{\pa u} -u \frac{\pa}{\pa v} + v_x \frac{\pa}{\pa u_x} -u_x \frac{\pa}{\pa v_x} + v_{xx} \frac{\pa}{\pa u_{xx}}  - u_{xx} \frac{\pa}{\pa v_{xx}} \ . \end{eqnarray*}
According to Lemma 3, these should be equivalent to a set of
standardly prolonged vector fields $Z_i$, given by $Z_i = A_i^{\
j} Y_j$; more precisely the latter should be the prolongation of
the vector fields $W_i = A_i^{\ j} Y_j$, and $A$ is determined by
\eqref{eq:L11}.

In this case we obtain
$$ A \ = \ \pmatrix{1 & u \cr 0 & 1 \cr} \ ; $$ this yields at once
\begin{eqnarray*}
W_1 &=& u (1+v) \frac{\pa}{\pa u} +  (v - u^2) \frac{\pa}{\pa v} \ , \\
W_2 &=& v \frac{\pa}{\pa u} -u \frac{\pa}{\pa v} \ ; \\
Z_1 &=& u (1+v) \frac{\pa}{\pa u} +  (v - u^2) \frac{\pa}{\pa v} + [u_x (1+ v) + u v_x] \frac{\pa}{\pa u_x} + (v_x - 2 u u_x) \frac{\pa}{\pa v_x} \\
 & & \ +  [u_{xx} (1 + v) + 2 u_x v_x + u v_{xx} \frac{\pa}{\pa u_{xx}} + (v_{xx} - 2 u u_{xx} - 2 u_x^2)  \frac{\pa}{\pa v_{xx}} \ , \\
Z_2 &=& v \frac{\pa}{\pa u} -u \frac{\pa}{\pa v} + v_x \frac{\pa}{\pa u_x} -u_x \frac{\pa}{\pa v_x} + v_{xx} \frac{\pa}{\pa u_{xx}}  - u_{xx} \frac{\pa}{\pa v_{xx}} \ . \end{eqnarray*}
It is immediate to check that $Z_i$ is the second standard prolongation of $W_i$.

\bigskip\noindent
{\bf Example 4 (non vertical vector fields).} The examples considered so far dealt with vertical vector fields; here we present, in the context of $\mu$-symmetries, an example where the vector field is not vertical. We will again deal with one independent ($x$) and two dependent ($u,v$) variables and second order prolongations; and use $\La$ -- and hence $A$ -- as in Example 2 above, see \eqref{eq:EX2}. We consider the vector field
\beq \label{eq:ex4VF} X \ = \ - \, \pa_x \ + \ u \, \pa_u \ + \ v \, \pa_v \ ; \eeq
its evolutionary representative is
\beq X_v \ = \ (u + u_x) \, \pa_u \ + \ (v + v_x) \, \pa_v \ . \eeq
Note moreover that the $X$-invariant functions are written as
\beq \label{eq:ex4IF} u \ = \ k_1 \, e^{-x}  \ , \ \ v \ = \ k_2 \, e^{-x} \ . \eeq
Applying $A$ as in \eqref{eq:EX2} on these, we obtain
\begin{eqnarray*}
W \ = \ A \cdot X &=& - \, \pa_x \ + \ u (1+v) \, \pa_u \ + \ v \, \pa_v \ , \\
W_v \ = \ A \cdot X_v &=& [u_x + u (1 + v + v_x)] \, \pa_u \ + \  (v + v_x) \, \pa_v \ . \end{eqnarray*}

The second order $\mu$-prolongation of $X$ and of $X_v$ are now easily computed (we will not write down the explicit formulas, which are rather long and of no special interest); we will also denote them by $Y = X_\mu^{(2)}$ and $Y_v = (X_v)_\mu^{(2)}$. Applying $A$ on these, we obtain $Z = A \cdot Y$ and $Z_v = A \cdot Y_v$.

Our Lemma 1, which requires to deal with vertical vector fields, states that $Z_v = W_v^{(2)}$, and this is indeed the case. On the other hand one can check that $Z \not= W^{(2)}$; more precisely, we have $Z = W^{(2)} + u_x v_x \pa_{u_x} + u_x v_{xx} \pa_{u_{xx}}$: this shows that the equivalence does indeed holds only between the evolutionary representatives. We stress here that $W_v = A \cdot X_v$ is \emph{not} the evolutionary representative of $W = A \cdot X$; the latter is
$$ W_{ev} \ =  \ [u_x + u (1+v)] \, \pa_u \ + \ (v + v_x) \, \pa_v \ . $$

Let us now consider, beside $Y,Y_v,Z,Z_v$ defined above, also the second standard prolongation of $X$ and $X_v$, i.e. $X^{(2)}_0$ and $(X_v)_0^{(2)}$; these have of course no special relation with $Z$, $W^{(2)}$ or $Z_v = W_v^{(2)}$. According to Lemma 2, however, we should have that $Y_v = Z_v = X_v^{(2)} $ (and actually vanish) on the set of invariant functions, and this is indeed the case.


\section{Conclusions}
\label{sec:conclusions}

In this paper we have considered known notions and results, and
provided a new way to look at them and interpretation with the aim
of advancing our understanding of twisted prolongations and of
their effectiveness in determining solutions to differential
equations.

We have considered the classical theory of symmetries of
differential equations; we have noted that what matters in
establishing if a vector field (an involution set of vector
fields) is a symmetry (a symmetry algebra) of a given differential
equation of order $n$ is not the prolongation of a vector field
(of a set of vector fields), but the distribution generated by
this (by these) in $J^n M$, so that changing the generators of the
distribution does not alter the property of being a symmetry (a
symmetry algebra).

We have then noted that changing the set of generators of the
above mentioned distribution still requiring that they are
projectable to each $J^k M$ (for $k \le n$), and for the setting
of ODEs that they satisfy the IBDP leads us to consider
$\lambda$-prolongations or, in the case of sets of vector fields,
$\sigma$-prolongations. On the other hand, requiring that the set
of invariant functions is the same as for the original generators
-- which is the natural requirement on the setting of PDEs, leads
us to consider $\mu$-prolongations. For sets of vector fields,
these can be combined with $\sigma$ ones to produce combined
twisted prolongations, i.e. $\chi$-prolongations.

The change of generators for the distribution is naturally
interpreted in terms of a gauge action. This holds both for $\la$
and $\mu$ prolongations, in which case the gauge action is on the
space of dependent variables and their variables; and in the case
of $\s$-prolongations, in which case it is on the space of vector
fields (see above for more details).

We have also stressed that while changing the set of generators
produces twisted symmetries, the correspondence between the gauge
action and the twisted symmetries is such that if we consider a
given twisted symmetry and look for a change of generators for the
distribution (this is naturally indexed by a gauge transformation)
corresponding to this, we will in general obtain a nonlocal map.

We conclude that while our discussion shows that twisted
symmetries can be understood in terms of change of generators for
the relevant distributions of prolonged vector fields, this point
of view would in general lead to consider non-local maps. In other
words, twisted symmetries are actually more general than standard
symmetries.

On the other hand, the general -- and generally formal, see above
-- correspondence unveiled by the present work simply explains why
twisted symmetries are as effective as standard ones in the
symmetry analysis of differential equations.


\section*{Appendix A. \\ Prolongation of vector fields in involution}

In this Appendix we establish the relation between the involution
relations for vector fields in $M$ and for their prolongation in
$J^n M$.

In other words, we want to consider vector fields of the form
$$ Z \ = \ f^\a (x,u) \ X_\a $$ and describe prolongations of $Z$
in terms of the prolongations of the $X_\a$, for general smooth
functions $f^\a$.

We will write the basis vector fields as $X_\a = \xi_\a^i \pa_i +
\phi_\a^a \pa_a$, and \beq\label{eq:B_Z} Z \ = \ f^\a \ X_\a \ = \
\Xi^i \, \pa_i \ + \ \Phi^a \, \pa_a \ ; \eeq needless to say, we
have $\Xi^i = f^\a \xi_\a^i$, $\Phi^a = f^\a \phi_\a^a$.

We will write the prolonged field as \beq\label{eq_B_Zn} Z^{(n)} \
= \ \Xi^i \, \pa_i \ + \ \chi^a_J \, \pa_a^J \eeq (sum over $J$ is
meant with $|J| \le n$), with $\chi^a_0 = \Phi^a$. We like to
write this prolongation as \beq\label{eq_B_Zn2} Z^{(n)} \ = \ f^\a
\, X_\a^{(n)} \ + \ \Ga^{(n)} \ . \eeq In terms of the
coefficients of the vector field $Z^{(n)}$, this means
\beq\label{eq:B_Psi} \chi^a_J \ = \ f^\a \, \psi^a_{\a;J} \ + \
W_{J}^a \ , \eeq where $\psi^a_{\a;J}$ are the coefficients in the
(standard) prolongation of $X_\a$.

We will also use the notation \beq \label{eq:AQ} Q^a_{\a;J} \ := \
\psi^a_{\a ; J} - u^a_{J,k} \xi^k_\a \ . \eeq

\medskip\noindent
{\bf Lemma A.1.} {\it The coefficients $W^a_J$ introduced above
satisfy the recursion relation \beq \label{eq:LA1} W_{J,i}^a \ = \
(D_i W_{J}^a ) \ + \ (D_i f^\a ) \ Q^a_{\a;J} \ , \eeq where
$W^a_0 = 0$.}

\medskip\noindent
{\bf Proof.} Applying the prolongation formula, we have
$$ \chi^a_{J,i} \ = \ D_i \chi^a_J \ - \ \( u^a_{J,k} \, D_i \Xi^k \) \ ; $$
recalling now the definition of the $\Xi$, see \eqref{eq:B_Z} and
the above expression \eqref{eq:B_Psi} for the $\chi$, this yields
\begin{eqnarray*}
\chi^a_{J,i} &=& D_i \( f^\a \psi^a_{\a;J} \, + \, W_{J}^a \) \ - \ u^a_{J,k} \, D_i \( f^\a \xi_\a^k \) \\
&=& (D_i f^\a ) \, [\psi^a_{\a ; J} - u^a_{J,k} \xi_\a^k ] \ + \ (D_i W_{J}^a ) \ + \
f^\a \, \( D_i \psi^a_{\a;J} \, - \, u^a_J D_i \xi^k_\a \) \\
&=& \psi^a_{\a; J,i} \ + \ (D_i W_{J}^a ) \ + \ (D_i f^\a ) \,
[\psi^a_{\a ; J} - u^a_{J,k} \xi^k_\a ] \ . \end{eqnarray*}

Using the notation \eqref{eq:AQ}, and going back to the expression
\eqref{eq:B_Psi}, the recursion relation for the $W$ results to be
\eqref{eq:LA1}. \EOP
\bigskip

This Lemma allows to give the generalization of the standard result about prolongations of Lie algebras of vector fields to the case of vector fields in
involution.

\medskip\noindent
{\bf Lemma A.2.} {\it Let the vector fields $X_\a$ ($\a =
1,...,r$) be in involution, $$ [X_\a , X_\b] \ = \ F_{\a \b}^\ga
(x,u) \ X_\ga \ .
$$ Then their prolongations satisfy \beq \label{eq:LA2a}
[X_\a^{(n)} , X_\b^{(n)} ] \ = \ F_{\a \b}^\ga \ X_\a^{(n)} \ + \
\Ga_{\a \b}^{(n)} \ , \eeq where the coefficients $W_{\a \b; J}^a$
of the vector fields $\Ga_{\a \b}^{(n)} = W_{\a \b; J}^a \pa_a^J $
satisfy $W^a_{\a \b ;0} = 0$ and the recursion relations \beq
\label{eq:LA2b} W_{\a \b;J,i}^a \ = \ (D_i W_{\a \b;J}^a ) \ + \
(D_i F_{\a \b}^\ga ) \ Q^a_{\ga;J} \ , \eeq with $W^a_{\a \b ;0} =
0$.}

\medskip\noindent
{\bf Proof.} Basically this follows by Lemma A.1, introducing the
two indices $\a$ and $\b$ in our previous discussion. We know by
Lemma 1 that
$$ \[ X_\a^{(n)} , X_\b^{(n)} \] \ = \ \( F_{\a \b}^\ga (x,u) \
X_\ga\)^{(n)} \ := \ Z_{\a \b}^{(n)} \ . $$ The vector field
$Z_{\a \b}$ on the r.h.s. is of the form considered in the
discussion leading to Lemma A.1; thus we know that
$$ Z_{\a \b}^{(n)} \ = \ F_{\a \b}^\ga \, X_\ga^{(n)} \ + \
\Ga_{\a \b}^{(n)} $$ where the vector fields $\Ga_{\a \b}^{(n)}$
can be written as $\Ga_{\a \b}^{(n)} = W_{\a \b; J}^a  \pa_a^J $.
Again by Lemma A.1, the coefficients $W_{\a \b; J}^a$ satisfy
\eqref{eq:LA2b}. \EOP


\section*{Appendix B. \\ Equivalent symmetries}

Our discussion led us to consider deformations of the prolongation operation.
It is thus advisable, when considering symmetries of a differential equation (say of order $n$), to keep track not only of the vector field $X$ in $M$, but also of the prolongation operation $\mathcal{P}$ used to generate from $X$ a vector field in $J^n M$. In this context we will denote by $(X,\P )$ the pair consisting of a vector field and a prolongation operation; we will denote by $\P_0$ the standard prolongation operation. When $\P \not= \P_0$ we have a twisted prolongation.
We will also denote by $X^{(n)}_\P$ the vector field in $J^n M$ obtained by prolonging $X$ via the $\P$ prolongation operation.

We can then generalize Definition 2 by the following (which reduces to Definition 2 for $\P = \P_0$)

\medskip\noindent
{\bf Definition B.1.} {\it The vector field $X$ in $M$ with the prolongation operation $\P$ is a \emph{$\P$-symmetry
generator} for the differential equation $\De$ of order $n$ if
$S_\De \ss J^n M$ is an integral manifold for the distribution
$\D_X$ spanned by $X^{(n)}_\P$ in $J^n M$.  The set $\X = \{ X_1 ,
... X_r \}$ of vector fields in $M$ with the prolongation operation $\P$
is a \emph{set of $\P$-symmetry generators} for the differential equation $\De$ of order $n$ if $S_\De \ss J^n M$ is an integral manifold for the distribution
$\D_\X^{(\P)}$ spanned by $X_{\a,\P}^{(n)}$ in $J^n M$.}
\bigskip

When we want to keep track of the prolongation used, we will denote the distribution by $\D_X^\P$, or the like for sets of vector fields.

As we consider general prolongation operations, it makes sense to consider vector fields which are prolonged through \emph{different} prolongation operations to produce the \emph{same} distribution in $J^n M$.

\medskip\noindent
{\bf Definition B.2.} {\it Assume the vector fields $X$ and $Y$ in $M$
are respectively a $\P$-symmetry and a $\P'$-symmetry generator of the differential equation $\De$ of order $n$, with $\P$ and $\P'$ possibly different. We say that $(X,\P)$ and $(Y,\P')$ are \emph{equivalent symmetries of $\De$} if they give raise to the same distribution in $J^n M$, i.e. if $\D_X = \D_Y$. The sets of vector fields $\X$ and $\Y$ in $M$, both symmetry sets of the
differential equation $\De$ of order $n$, are \emph{equivalent sets
of symmetries of $\De$} if they span the same distribution in $J^n
M$, i.e. if $\D_\X^{(\P)} = \D_\Y^{(\P')}$.}
\bigskip

We stress that in the case of sets, equivalent sets are not required to have the same cardinality: indeed, we have not required that the $X_\a$ (and the $Y_\a$), nor the $X_{\a,\P}^{(n)}$ (and the $Y_{\a,\P'}^{(n)}$) be independent at each point, and not even that they are sets of independent vector fields.

The results of Section \ref{sec:distrib} can be restated in terms of equivalent symmetries. In particular, Corollary 1 guarantees that $(X,\P_\la)$ and $(\wt{X},\P_0)$ are equivalent symmetries; similarly, Corollary 3 guarantees that the sets $(\X,\P_\s )$ and $(\mathcal{W},\P_0)$ are equivalent sets.

\vfill

{ } \hfill {\tt \giorno}


\begin{thebibliography}{99}


{\small

\bibitem{EMS1} D.V. Alekseevsky, A.M. Vinogradov and V.V. Lychagin,
{\it Basic ideas and concepts of differential geometry}, Springer
1991

\bibitem{BP} A. Barco and G.E. Prince, ``Solvable symmetry structures in
differential form'', {\it Acta Appl. Math.} {\bf 66} (2001),
89-121

\bibitem{BH} P. Basarab-Horwath,  ``Integrability by quadratures for
systems of involutive vector fields'', {\it Ukr. Math. J.} {\bf
43} (1992), 1236-1242

\bibitem{Bir} G. Birkhoff, {\it Hydrodynamics. A study in logic, fact,
and similitude}, Princeton 1950

\bibitem{BCM} Ph. Broadbridge, C. Chanu and W. Miller,
``Solutions of Helmholtz and Schroedinger equations with side condition and nonregular separation of variables'', {\it SIGMA} {\bf 8} (2012), 089

\bibitem{CDW} J.F. Carinena, M. Del Olmo and P. Winternitz,
``On the relation between weak and strong invariance of
differential equations'', {\it Lett. Math. Phys.} {\bf 29} (1993),
151-163

\bibitem{CFM1} D. Catalano-Ferraioli and P. Morando,
``Local and nonlocal solvable structures in the reduction of
ODEs'', {\it J. Phys. A: Math. Theor.} {\bf 42} (2009), 035210

\bibitem{CFM2} D. Catalano-Ferraioli and P. Morando,
``Applications of solvable structures to the nonlocal symmetry
reduction of ODEs'', {\it J. Nonlin. Math. Phys.} {\bf 16-S}
(2009), 27-42

\bibitem{CFM3} D. Catalano-Ferraioli and P. Morando,
``Nonlocal interpretation of $\la$-variational symmetry-reduction
method'', {\it arXiv:0903.1014} (2009)

\bibitem{CFX} D. Catalano-Ferraioli and L.A. de Oliveira Silva,
``Nontrivial 1-parameter families of zero-curvature
representations via symmetry actions'', preprint 2014

\bibitem{CCL} S.S. Chern, K.S. Lam and W.H. Chen, {\it Lectures on differential geometry}, World Scientific 1999.

\bibitem{Cds1} G. Cicogna, ``Reduction of systems of first-order
differential equations via $\Lambda$-symmetries'', {\it Phys.
Lett. A} {\bf 372} (2008), 3672-3677

\bibitem{Cds2} G. Cicogna, ``Symmetries of Hamiltonian
equations and $\Lambda$-constants of motion'', {\it  J. Nonlin.
Math. Phys.} {\bf 16} (2009), 43-60

\bibitem{CGbook} G. Cicogna and G. Gaeta,
{\it Symmetry and perturbation theory in nonlinear dynamics},
Springer 1999

\bibitem{CGpart} G. Cicogna and G. Gaeta, ``Partial Lie-point symmetries
of differential equations'', {\it J. Phys. A} {\bf 34} (2001),
491-512



\bibitem{CGMor} G. Cicogna, G. Gaeta and P. Morando, ``On the relation
between standard and $\mu$-symmetries for PDEs'', {\it J. Phys. A}
{\bf 37} (2004), 9467-9486

\bibitem{Sprol}  G. Cicogna, G. Gaeta and S. Walcher, ``A generalization
of $\la$-symmetry reduction for systems of ODEs:
$\s$-symmetries'', {\it J. Phys. A: Math. Theor.} {\bf 45} (2012)
355205 (29 pp)

\bibitem{SprolDS}  G. Cicogna, G. Gaeta and S. Walcher: ``Dynamical
systems and $\sigma$-symmetries'', {\it J. Phys. A} {\bf 46}
(2013), 235204 (23pp)

\bibitem{CGWjlt1}  G. Cicogna, G. Gaeta and S. Walcher, ``Orbital reducibility
and a generalization of lambda symmetries'', {\it J. Lie Theory}
{\bf 23} (2013), 357-381

\bibitem{CGWjlt2}  G. Cicogna, G. Gaeta and S. Walcher, ``Side conditions
for ordinary differential equations'', preprint 2014, to appear in
{\it J. Lie Theory} {\bf 25} (2015), 125-146

\bibitem{ClK} P.A. Clarkson and M.D. Kruskal, ``New similarity reductions
of the Boussinesq equation'', {\it J. Math. Phys.} {\bf 30}
(1989), 2201-2213

\bibitem{FT} L.D. Faddeev and L.A. Takhtajan,
{\it Hamiltonian methods in the theory of solitons}, Springer 1987

\bibitem{Gaebook} G. Gaeta, {\it Nonlinear symmetries and nonlinear
equations}, Kluwer, Dordrecht 1994


\bibitem{GJGP} G. Gaeta, ``A gauge-theoretic description of $\mu$-prolongations,
and $\mu$-symmetries of differential equations'', {\it J. Geom. Phys.} {\bf 59}  (2009), 519-539


\bibitem{Gtwist1} G. Gaeta, ``Twisted symmetries of differential
equations'', {\it J. Nonlin. Math. Phys.}, {\bf 16-S} (2009),
107-136

\bibitem{GJPAgau} G. Gaeta, ``Gauge fixing and twisted
prolongations'', {\it J. Phys. A} {\bf 44} (2011), 325203 (9pp)

\bibitem{Gtwist2} G. Gaeta, ``Simple and collective twisted symmetries'',
preprint 2014, to appear in {\it J. Nonlin. Math. Phys.} {\bf 21} (2014), 593-627

\bibitem{GMor} G. Gaeta and P. Morando, ``On the geometry of lambda-symmetries and PDE reduction'', {\it J. Phys. A} {\bf 37} (2004), 6955-6975

\bibitem{God} C. Godbillon, {\it G\'eom\'etrie Diff\'erentielle et M\'ecanique Analitique}, Hermann 1969

\bibitem{HaWa} K.P. Hadeler and S. Walcher, ``Reducible ordinary differential equations'', {\it J. Nonlinear Sci.} {\bf 16} (2006), 583-613

\bibitem{HeW1} W. Hereman, ``Review of symbolic software for Lie symmetry analysis'', {\it Math. Comp. Mod.} {\bf 25} (1997), 115-132

\bibitem{HeW2} W. Hereman, ``Review of symbolic software for the computation of Lie symmetries of differential equations'', {\it Euromath Bull.} {\bf 1} (1994), 45-82

\bibitem{HeW3} W. Hereman, ``Symbolic computation of conservation laws of nonlinear partial differential equations in multi-dimensions'', {\it Int. J. Quantum Chem.} {\bf 106} (2006), 278-299

\bibitem{Her} R. Hermann, {\it Differential Geometry and the Calculus of
Variations}, Academic Press (New York) 1968

\bibitem{KrV} I.S. Krasil'schik and A.M. Vinogradov,
{\it Symmetries and conservation laws for differential equations
of mathematical physics}, A.M.S. 1999

\bibitem{LeWi1} D. Levi and P. Winternitz, Non-classical symmetry reduction:
example of the Boussinesq equation'', {\it J. Phys. A: Math. Gen.}
{\bf 22} (1989), 2915-2924

\bibitem{LeWi2} D. Levi and P. Winternitz, ``Symmetries and conditional symmetries
of differential-difference equations'', {\it J. Math. Phys.} {\bf 34} (1993), 3713-3730

\bibitem{Mar} M. Marvan, ``On zero curvature representations of partial differential equations'', pp. 103-122 in: {\it Differential Geometry and Applications}, Silesian University Opava, 1993; online at
    {\tt http://www.emis.de/proceedings/5ICDGA}

\bibitem{Mordef} P. Morando, ``Deformation of Lie derivative and $\mu$-symmetries'', {\it J. Phys. A} {\bf 40} (2007), 11547-11560


\bibitem{MuRom1} C. Muriel and J.L. Romero, ``New methods of reduction for ordinary differential equations'', {\it IMA J. Appl. Math.} {\bf 66} (2001), 111-125

\bibitem{MuRom2} C. Muriel and J.L. Romero, ``$C^\infty$ symmetries and nonsolvable symmetry algebras'', {\it IMA J. Appl. Math.} {\bf 66} (2001), 477-498

\bibitem{MuRom3} C. Muriel and J.L. Romero, ``Prolongations of vector fields and the invariants-by-derivation property'', {\it Theor. Math. Phys.} {\bf 113} (2002), 1565-1575

\bibitem{MuRom4} C. Muriel and J.L. Romero, ``$C^\infty$-symmetries and integrability of ordinary differential equations''; pp. 143-150 in {\it Proceedings of the I Colloquium on Lie theory and applications (Vigo)}, 2002

\bibitem{MuRom5}  C. Muriel and J.L. Romero, ``$C^\infty$-symmetries and nonlocal symmetries of exponential type, {\it IMA J. Appl. Math.} {\bf 72} (2007) 191-205

\bibitem{MuRom6}  C. Muriel and J.L. Romero, ``Integrating factors
and lambda-symmetries'', {\it J. Nonlin. Math. Phys.} {\bf 15- S3}
(2008), 300-309

\bibitem{MuRom7} C. Muriel and J.L. Romero, ``First integrals,
integrating factors and $\lambda$-symmetries of second-order
differential equations'', {\it J. Phys. A} {\bf 42} (2009), 365207

\bibitem{MuRom8} C. Muriel and J.L. Romero, ``Nonlocal symmetries, telescopic
vector fields and $\la$-symmetries of ordinary differential
equations'', {\it SIGMA} {\bf 8} (2012), 106

\bibitem{MRO} C. Muriel, J.L. Romero and P.J. Olver, ``Variational
$C^\infty$ symmetries and Euler-Lagrange equations'', {\it J.
Diff. Eqs.} {\bf 222} (2006), 164-184

\bibitem{NuL} M.C. Nucci and P.G.L. Leach, ``The determination of nonlocal symmetries
by the technique of reduction of order'', {\it J. Math. Anal.
Appl.} {\bf 251} (2000), 871-884

\bibitem{Oli} F. Oliveri, ``Lie symmetries of differential
equations: classical results and recent contributions'', {\it
Symmetry} {\bf 2} (2010), 658-706

\bibitem{Olv1} P.J. Olver,
{\it Application of Lie groups to differential equations},
Springer 1986

\bibitem{Olv2} P.J. Olver,
{\it Equivalence, Invariants and Symmetry}, Cambridge University
Press 1995

\bibitem{OR1} P.J. Olver and Ph. Rosenau, ``The construction of special solutions to partial differential equations'', {\it Phys. Lett. A} {\bf 114} (1986), 107-112

\bibitem{OR2} P.J. Olver and Ph. Rosenau, ``Group-invariant solutions of differential equations'', {\it SIAM J. Appl. Math.} {\bf 47} (1987), 263-278

\bibitem{Ovs} L.V. Ovsjannikov, {\it Group Analysis of Differential Equations}
(in Russian), Nauka 1962

\bibitem{PuS} E. Pucci and G. Saccomandi, ``On the reduction methods
for ordinary differential equations'', {\it J. Phys. A} {\bf 35}
(2002), 6145-6155

\bibitem{Sch} F. Schwarz, ``Symmetries of differential equations: from Sophus Lie
to computer algebra'', {\it SIAM Rev.} {\bf 30} (1988), 450-481

\bibitem{Sha} R.W. Sharpe, {\it Differential Geometry}, Springer 1997

\bibitem{ShP} J. Sherring and G. Prince, ``Geometric aspects of reduction of order'',
{\it Trans. Am. Math. Soc.} {\bf 334} (1992), 433-453

\bibitem{Ste} H. Stephani,
{\it Differential equations. Their solution using symmetries},
Cambridge University Press 1989

\bibitem{Stern} S. Sternberg {\it Lectures on Differential Geometry}, Chelsea 1983

\bibitem{Wal99} S. Walcher, ``Multi-parameter symmetries of first
order ordinary differential equations'',  {\it J. Lie Theory} {\bf
9} (1999), 249-269

\bibitem{WalSPT} S. Walcher, ``Orbital symmetries of first order ODEs'', in {\it
Symmetry and Perturbation Theory (SPT98)} (pp. 96-113), Editors: A. Degasperis
and G. Gaeta, World Scientific 1999

}

\end{thebibliography}
\end{document}